\documentclass[a4paper]{article} 
\usepackage{jheppub}
\usepackage{slashed}
\usepackage{multirow}
\usepackage{multicol}

\newcommand{\comCL}[1]{{\color{green}\textsf{\small [#1]}}}
\newcommand{\mttwo}{M_{T2}}
\newcommand{\hide}[1]{}

\newcommand{\rhsFourQISMinusOne}{{\chi^2 
    +A_T + 
      \sqrt{\left(A_T^2 - m_a^2 m_b^2 \right)\left(
       1 + \frac{4 \chi^2}{2 A_T - m_a^2 - m_b^2}\right)}}}
\newcommand{\rhsOne}{{\rhsFourQISMinusOne}}

\newcommand{\rhsDegenMassArbQ}{{
\chi^2 + (1+Q)m^2
    -A_T Q + 
      \sqrt{\left(A_T + m^2 \right)
\left(Q^2 \left(A_T-m^2\right)+ 2 \chi^2 \right)}
}}

\newcommand\singleExpression{{\chi^2 + (1+Q)\frac{m_a^2  + m_b^2}{2}
    -A_T Q + 
      \sqrt{\left(A_T^2 - m_a^2 m_b^2 \right)\left(
       Q^2 + \frac{4 \chi^2}{2 A_T - m_a^2 - m_b^2}\right)}
}}

\newcommand{\rhsThree}{{
\begin{cases}
M_{T2}^2(a^\mu,b^\mu,\slashed{\bf p}_T=0) \qquad\qquad\text{(see equation~(\ref{eq:mt2ptmisszero}))} & \text{if $Q=0$} \\
\rhsFourQISMinusOne & \text{if $Q=-1$}\\ 
\rhsDegenMassArbQ & \text{if $m_a=m_b=m$}\\
\text{use numerical methods} & \text{otherwise.}
\end{cases}
}}
\newcommand{\rhsThreeAlt}{{
\begin{cases}
M_{T2}^2(a^\mu,b^\mu,\slashed{\bf p}_T=0) \qquad\qquad\text{(see equation~(\ref{eq:mt2ptmisszero}))} & \text{if $Q=0$} \\
\singleExpression & \text{if $Q=-1$ or $m_a=m_b$}\\ 
\text{use numerical methods} & \text{otherwise.}
\end{cases}
}}

\newcommand{\rhsForZeroPtmissPartOne}{
\chi^2 + A_T
-\frac{ 2 (A_T -m_a^2)(A_T-m_b^2)}{2 A_T - m_a^2 - m_b^2}
}
\newcommand{\rhsForZeroPtmissPartOneAlternative}{
\chi^2 + \frac{m_a^2+m_b^2}{2}+\frac{(m_b^2-m_a^2)^2}{2(2A_T-m_a^2-m_b^2)}
}
\newcommand{\rhsForZeroPtmissPartTwo}{
\sqrt{(A_T^2-m_a^2 m_b^2)\left(1+\frac{4 \chi^2}{2 A_T - m_a^2 - m_b^2}-\frac{4(A_T -m_a^2)(A_T-m_b^2)}{(2 A_T - m_a^2 - m_b^2)^2}\right)}
}
\newcommand{\rhsForZeroPtmissPartTwoAlternative}{
\sqrt{(A_T^2-m_a^2 m_b^2) \left( \left(\frac{m_b^2-m_a^2}{2A_T-m_a^2-m_b^2}\right)^2 +\frac{4 \chi^2}{2 A_T - m_a^2 - m_b^2}\right)}
}

\newcommand{\eap}{{  ( \hat{\bf a} . {   \hat{\slashed{\bf p}}   }' )  }}
\newcommand{\ebp}{{  ( \hat{\bf b} . {   \hat{\slashed{\bf p}}   }' )  }}
\newcommand{\esigmap}{{  ( {\bf \sigma} . {   \hat{\slashed{\bf p}} }' )  }}
\newcommand{\edeltap}{{  ( {\bf \delta} . {   \hat{\slashed{\bf p}} }' )  }}
\newcommand{\sigmadotp}{{({\bf \sigma}.{ \hat{\slashed{\bf p}}}) }}
\newcommand{\deltadotp}{{({\bf \delta}.{ \hat{\slashed{\bf p}}}) }}
\newcommand{\ahdotph}{{(\hat{\bf a}.{ \hat{\slashed{\bf p}}}) }}
\newcommand{\bhdotph}{{(\hat{\bf b}.{ \hat{\slashed{\bf p}}}) }}
\newcommand{\adotpFull}{{({\bf a}.{ {\slashed{\bf p}}}) }}

\newcommand{\acrospFull}{{({\bf a}\wedge{ {\slashed{\bf p}}}) }}
\newcommand{\bcrospFull}{{({\bf b}\wedge{ {\slashed{\bf p}}}) }}
\newcommand{\adotp}{c_{ap}}
\newcommand{\bdotp}{c_{bp}}
\newcommand{\acrosp}{s_{ap}}
\newcommand{\bcrosp}{s_{bp}}
\newcommand{\eahbh}{{   ({\bf \hat  a}.{{\bf \hat b}'})    }}
\newcommand{\ahdotbh}{{   ({\bf \hat a}.{{\bf \hat  b}})    }}
\newcommand{\cosTheta}{{\cos\theta}}
\newcommand{\sinTheta}{{\sin\theta}}
\newcommand{\poverq}{{\rho}}
\newcommand{\vecPtmiss}{{\slashed{\bf p}}}
\newcommand{\magPtmiss}{{|{\slashed{\bf p}}|}}
\newcommand{\thing}{{K}}
\newcommand{\LHS}{{L_1}}
\newcommand{\RHS}{{L_2}}
\newcommand{\Kone}{{K_1}}
\newcommand{\Ks}{{K_s}}
\newcommand{\Kc}{{K_c}}
\newcommand{\Kss}{{K_{ss}}}
\newcommand{\Kcc}{{K_{cc}}}
\newcommand{\Kcs}{{K_{cs}}}
\newcommand{\AAA}{{A}}
\newcommand{\BB}{{B}}
\newcommand{\CC}{{C}}
\newcommand{\DD}{{D}}
\newcommand{\EE}{{E}}
\newcommand{\agoestob}{{\left[ a \leftrightarrow b \right]}}

\begin{document}
\title{The stransverse mass, $\mttwo$, in special cases}
%
\author{Christopher~G.~Lester}
\affiliation[a]{University of Cambridge,\\Department of Physics,\\Cavendish Laboratory,\\JJ Thomson Avenue,\\ Cambridge,\\United Kingdom}
\emailAdd{lester@hep.phy.cam.ac.uk}

\abstract{
This document describes some special cases in which the stransverse mass, $\mttwo$, may be calculated by non-iterative algorithms.  The most notable special case is that in which the visible particles and the hypothesised invisible particles are massless -- a situation relevant to its current usage in the Large Hadron Collider as a discovery variable, and a situation for which no analytic answer was previously known.  We also derive an expression for $\mttwo$ in another set of new (though arguably less interesting) special cases in which the missing transverse momentum must point parallel or anti parallel to the visible momentum sum.  In addition, we find new derivations for already known $\mttwo$ solutions in a manner that maintains manifest contralinear boost invariance throughout, providing new insights into old results.  Along the way, we stumble across some unexpected results and make conjectures relating to geometric forms of $M_\mathrm{eff}$ and $H_T$ and their relationship to $\mttwo$.
}
\arxivnumber{1103.5682}
\maketitle

\begin{multicols}{2}

\section{Introduction}

The purpose of this note is to show how $\mttwo$
\cite{Lester:1999tx} (also known as the ``stransverse'' mass) may be calculated by non-iterative algorithms in a small number of special cases, some of which are relevant to current usage patterns at the Large Hadron Collider.

There are at most ten people who will find this document interesting,
and all of them would skip past any ``motivating introductory waffle''
if it were to be supplied.  The remainder of the earth's population
will, quite rightly, find their local telephone directory a much
better bed-time read, no matter how much effort I put in to motivating
it.  Insomniacs, journal referees, or those who are shocked to the
core by the idea that a note may fail to motivate itself anywhere
other than in the abstract or conclusions, may find in the Appendix
some additional discussion of currency and relevance.  However here in
the introduction it would seem to make better sense to get straight
down to business.

In Section~\ref{sec:notation} we describe our notation. In
Section~\ref{sec:themassssslesscase} we find the first non-iterative
expression for $\mttwo$ valid for the case where all particles are
massless.  At the end of that section we interpret this result
and speculate on what it might be telling us.  In
Section~\ref{sec:funnycases} we start all over again, finding non-iterative expressions for $\mttwo$ which are valid when the missing transverse momentum points in special directions. Many of the sub cases in Section~\ref{sec:funnycases} are new, and even where they are not, insight is provided by the new derivations. 

\subsection*{On the applicability of the results herein.}

It was noted in \cite{Barr:2003rg,Lester:2007fq} that the input momenta supplied to $\mttwo$ may be grouped into two types -- those that lead to ``balanced'' solutions, and and those that lead to ``unbalanced'' solutions.  See \cite{Lester:2007fq} for details of how these are defined.   Both the general solution for $\mttwo$ for ``unbalanced'' inputs, and the test that may be applied to inputs to determine whether or not they are ``unbalanced'', are simple and have been known for some time.  They may be found in \cite{Lester:2007fq}.  In contrast, it is not expected that a general non-iterative algorithm for computing $\mttwo$ for inputs that lead to ``balanced'' configurations exists.  Indeed, from that statement derives part of the interest in the special cases considered herein for which non-iterative algorithms can be found.

Consequently, this note is only interested in determining and recording non-iterative solutions to MT2 for the case of inputs leading to ``balanced'' configurations.  The unbalanced solution is already known in all generality.  It will be assumed that all results in this note are taken to be ``preceded'' by a test for ``unbalancedness'' following \cite{Lester:2007fq}, and that the results herein are only to be applied if that test fails.

On a separate matter we\footnote{This was pointed out to me in the first instance by Chris Young.} note that\begin{equation}\mttwo(a^\mu,b^\mu,\slashed{p}) = M_{CT2}(a^\mu,b^\mu,-\slashed{p})\label{eq:mt2ismct2backwards}\end{equation} and so the results herein may trivially be turned into non-iterative expressions for $M_{CT2}$ \cite{Cho:2009ve} by changing the sign of the missing transverse momentum.

\section{Notation}
\label{sec:notation}

The stransverse mass, $\mttwo$, \cite{Lester:1999tx} is\footnote{See proof in \cite{Cheng:2008hk}} the maximal lower bound on the mass of each member of a pair of identical parent particles which, if pair-produced at a hadron collider, could have each undergone a two-body
decay into (i) a visible particle (or collection of particles) and (ii) an invisible object of hypothesised mass $\chi$.
The decay products of each parent are referred to as coming from
different ``sides'' of the event. The momenta of the visible decay
products of sides 1 and 2 will be referred to as $a^\mu$ and $b^\mu$
respectively.  The momenta of the invisible daughters of sides 1 and 2 (or
more usually the hypothesised momenta that they might take in some
part of a calculation, since the true momenta are unknown) are
referred to as $p^\mu$ and $q^\mu$ respectively.  A consequence of the
definition of $\mttwo$ (as a hadron collider variable) is that it is
insensitive to the $z$ components of the each of the visible momenta
$a^\mu$ and $b^\mu$. It is only sensitive to the masses and transverse momenta of those objects.  Accordingly, when performing
our calculations we work exclusively in a Minkowski space of dimension
3=1+2 with signature $(+,-,-)$ rather than the usual 4=1+3 dimensions
with signature $(+,-,-,-)$.\footnote{Accordingly, unless explicitly
stated otherwise, any references to ``Lorenz vectors'' or ``boosts'',
etc, must be assumed to be in the reduced transverse space with
signature $(+,-,-)$.}  Accordingly we view $a^\mu$ as containing three
components: $a^\mu=(e_T, p_x,p_y) = (e_T, {\bf p})$ where ${\bf
p}=(p_x,p_y)$ and where $p_x$ and $p_y$ are the transverse components
of $a$'s momentum, and $e_T$ is defined by $e_T =
\sqrt{m^2+p_x^2+p_y^2}$, in which $m$ is the the ``actual'' (1+3
dimensional) mass of the particle.  We denote the missing momentum
2-vector (another input to $\mttwo$) as $\vecPtmiss =
(\slashed{p}_x,\slashed{p}_y)$. We denote by $\chi$ the hypothesised
mass of the species of invisible particle that was generated in the
decay of each of the parents whose mass $\mttwo$ seeks to bound.

We use the usual Einstein summation convention and index raising and
lowering notation to allow us to construct objects which are scalars
with respect to ``intrinsic'' Lorentz transformations and rotations
within our reduced (+,-,-) transverse space.  For example, if
$a^\mu=(e_T, {\bf p})$ then $a_\mu=(e_T, -{\bf p})$ and $a^\mu a_\mu =
e_T^2 - |{\bf p}|^2$, and $a^\mu b_\mu$ is a scalar in our reduced
space.

Additionally, we will find it convenient to introduce an additional (non-standard) ``bar notation'' which is closely related to index raising and lowering. The ``bar'' creates a new Lorentz vector from an existing one, by reversal of the direction of its spatial component.  Thus if $a^\mu=(e_T, {\bf p})$ then ${\bar a}^\mu=(e_T, -{\bf p})$.  At first sight this may seem to be a backward step.  Have we not already introduced index lowering?  Surely ${\bar a}^\mu = a_\mu$, what do we need the bar's for?   Hopefully the utility will become apparent in use.  In short it is because ``scalars'' are not the only quantities we are interested in in the transverse plane, and we want the nature of our non-scalars to be evident without recourse to an abundance of indices.   When constructing simple expressions for $M_{T2}$ in special cases it is found often to be expedient to construct quantities which are not
representations of the the Lorentz group.  Specifically we want to be
able to construct quantities from pairs of Lorentz vectors which are
{\em not} Lorentz scalars.  One example is the quantity: $E_T^a E_T^b
+ {{\bf a}_T}.{{\bf b}_T}$.  Although this quantity (which is, for
historical reasons known as $A_T$\footnote{Note that the letter $A$ in
$A_T$ has no connection to the letter $a$ in $E_T^a$ or ${{\bf
a}_T}$.} and is closely related to $M_{CT}$ \cite{Tovey:2008ui}) is manifestly not invariant under Lorentz transformations,
it is nonetheless invariant under simultaneously applied boosts of equal
magnitude {\em but opposite direction} of the constituent vectors
$a^\mu$ and $b^\mu$.  The usefulness of such a quantity was first
noted in \cite{Tovey:2008ui}, wherein such transformations were named
``contralinear boosts'', and we can thus consider $A_T$ to be a
``contralinear boost invariant scalar'' or ``contra-scalar'' for short.  At first sight, we do not
appear to need to introduce any special notation to describe
quantities like $A_T$.  For example, the Einstein summation convention
itself allows us to write $A_T = a^\mu a^\mu$ or equivalently as $A_T
= a_\mu a_\mu$.  However, as we will need to work with both scalars and contra-scalars, sometimes in the same expression at the time time, the barred notation has the benefit of allowing us to suppress indices while retaining a clear understanding of whether a term represents a scalar ``$a.b$'' or a contra-scalar ``$a.{\bar b}$''.  To summarise the consequences of the ``bar'' notation:
\begin{itemize}
\item
$a$ differs from $\bar a$ only in the sign of the spatial components.  It exists to allow us to suppress indices in certain types of contractions while making the transformation properties of those contractions explicit.
\item
${\bar a}^\mu = a_\mu$ and ${\bar a}_\mu = a^\mu$.
\item
We suppress all contracted Lorentz indices, wherever possible, since they behave in the usual manner, i.e. $a.b=a^\mu b_\mu$ and $a.{\bar b}=a^\mu {\bar b}_\mu$ etc.
\item
Whereas $a.b$ and ${\bar a}.{\bar b}$ are Lorentz scalars, $a.{\bar b}$ and ${\bar a}.b$ are ``contralinear boost invariant scalars'' or ``contra-scalars'' for short.
\item
Trivially we have $a.b = {\bar a}.{\bar b}$ and $a.{\bar b} = {\bar a}.b$.
\end{itemize}


Finally, we define the ``Upstream Transverse Momentum'' (UTM) to be the physical transverse momentum against which the visible systems ${\bf a}$ and ${\bf b}$ and the missing transverse momentum are recoiling.   Denoting the transverse component of the UTM by ${\bf g}$ we have the relation ${\bf a}+{\bf b}+\vecPtmiss+{\bf g}={\bf 0}$ reminding us that these momenta are not all independent.

\par

\end{multicols}

\section{$M_{T2}$ in the fully massless case ($a^2=b^2=\chi^2=0$).}
\label{sec:themassssslesscase}

In this section it is our intention to write down a non-iterative expression for $\mttwo$ in the ``fully massless case'' i.e.~in the case in which the input particles and the invisible particles are taken to be massless $a^2=b^2=\chi^2=0$.  To the best of our knowledge this solution has not been reported elsewhere.
 This ``fully massless case'' is the situation in which $\mttwo$ is most frequently used, including cases such as the dijet configuration used in LHC supersymmetry searches \cite{Barr:2009wu,Barr:2010ii,Collaboration:1273174,Collaboration:2011qk}.\footnote{The fully massless case would not be appropriate where the visible momenta on each ``side'' are compound objects with significant masses -- such as when $\mttwo$ is used on dileptonic $t\bar t$ events to measure the top mass \cite{Cho:2008cu,Aaltonen:2009rm}.}

We begin by noting that if the vector $\vecPtmiss$ happens to lie ``between'' $\bf a$ and $\bf b$ (i.e. if $\vecPtmiss$ lies inside in the smaller of the two sectors of the transverse plane bounded by  $\bf a$ and $\bf b$) then $\mttwo$ in the fully massless case must be identically zero.  We call this a ``trivial zero'' of $\mttwo$ in the fully massless case.  One may prove that such a trivial zero exists because the constraint ${\bf p}+{\bf q}=\vecPtmiss$ can be solved by taking ${\bf p}\propto {\bf a}$ and  ${\bf q}\propto {\bf b}$. This is a partition of the missing transverse momentum that assigns transverse masses of zero to both sides of the event.  In the notation that will be introduced later in this section, it is straightforward to see that this ``trivial zero'' of $\mttwo$ in the fully massless case occurs when $\eap  \eahbh \ge 0$ and  $ \ebp  \eahbh \le 0$.

We now exclude this trivial zero, and instead consider what happens when $\vecPtmiss$ does {\em not} lie between $\bf a$ and $\bf b$.  Here we already know (see e.g.~equation (50) of \cite{Barr:2003rg}) that (i) that the splitting hypothesis which leads to the {\em minimal} ``balanced'' configuration satisfies the following relationship
in terms of ``transverse velocities'' ${\bf v}= {\bf p} /e_T $
\begin{equation}
({\bf v}_p - {\bf v}_a )\propto ({\bf v}_q - {\bf v}_b)\label{eq:vels}
\end{equation}
 where the proportional symbol means parallel, and (ii) that when both visible particles are massless, the splitting hypothesis which leads to the {\em minimal} $M_{T2}$ solution is a ``balanced'' configuration.
In the massless case, the transverse velocities are all represented by 2-D unit vectors since $e_T=|{\bf p}|$.  We will therefore re-write equation~(\ref{eq:vels}) as 
\begin{equation}
(\hat {\bf p} - \hat{\bf a} )\propto (\hat{\bf q} - \hat{\bf b}).\label{eq:vels2}
\end{equation}
A direct consequence of equation~(\ref{eq:vels2}) is that the angle {\em between} $q$ and $p$ at the MT2 solution is fixed to the same value as the angular separation between $a$ and $b$.\footnote{Note the ordering is $\theta_{qp} = - \theta_{pq} =  \theta_{ab} = -\theta_{ba}$.}   One possible such arrangement is therefore $\hat{\bf p}=\hat{\bf b}$ and $\hat {\bf q}=\hat {\bf a}$.  The general configuration allowed by equation~(\ref{eq:vels2}) can thus be parametrised by rotating this particular solution by an arbitrary angle $\theta$.  In other words, we can parametrise $\hat {\bf p}$ and $\hat {\bf q}$ in the following way:
\begin{eqnarray}
\hat {\bf p} &=& \cosTheta {\hat{\bf b}} + \sinTheta {\hat{\bf b}'}\label{eq:pparam} \\
\hat {\bf q} &=& \cosTheta {\hat{\bf a}} + \sinTheta {\hat{\bf a}'}\label{eq:qparam}
\end{eqnarray}
where we intend the two vector ${\hat{\bf a}'}$ to be obtained from the two vector ${\hat{\bf a}}$ by a rotation by +90 degrees in the transverse plane, and likewise ${\hat{\bf b}'}$ to be obtained from ${\hat{\bf b}}$ by the same rotation.
In effect, all that remains to do is to find $\theta$, $\left|{\bf p}\right|$ and  $\left|{\bf q}\right|$ by imposing the remaining constraints, namely (i) that the configuration be ``balanced'', i.e.
\begin{equation}
2\left(\left|{\bf a}\right|\left|{\bf p}\right|-{\bf a}.{\bf p}\right)
=
2\left(\left|{\bf b}\right|\left|{\bf q}\right|-{\bf b}.{\bf q}\right)\label{eq:balconfigsdfsdf}
\end{equation}
and (ii) that the momentum splitting condition is satisfied:
\begin{equation}
{\bf p} + {\bf q} = \slashed{\bf p}.\label{eq:splittingcondpoiy}
\end{equation}
Substitution of the parametrisation of (\ref{eq:pparam}) and (\ref{eq:qparam}) into (\ref{eq:balconfigsdfsdf}) leads to the constraint:
\begin{equation}
\left|{\bf a}\right|\left|{\bf p}\right|\left(1-\cosTheta \ahdotbh - \sinTheta \eahbh\right)
=\label{eq:fgdfguytr}
\left|{\bf b}\right|\left|{\bf q}\right|\left(1-\cosTheta \ahdotbh + \sinTheta \eahbh\right)
\end{equation}
while substitution into the splitting condition of (\ref{eq:splittingcondpoiy}) and taking the dot-product with ${\hat{\slashed{\bf p}} }'$ (a unit-two vector obtained by rotating  $\hat{\slashed{\bf p}}$ by +90 degrees in the transverse plane) leads to the constraint:
\begin{equation}
+\left|{\bf p}\right|\left(\cosTheta \ebp + \sinTheta \bhdotph\right)
= \label{eq:mombalone}
-\left|{\bf q}\right|\left(\cosTheta \eap + \sinTheta \ahdotph\right).
\end{equation}
All dependence on $\left|{\bf p}\right|$ and $\left|{\bf q}\right|$ may then be eliminated by taking the quotient of the last two constraints (\ref{eq:fgdfguytr}) and  (\ref{eq:mombalone}), which results in a single constraint of the form
\begin{equation}
\Kss\sin^2\theta + \Kcc \cos^2\theta + \Kcs \cos \theta \sin \theta + \Ks \sin\theta + \Kc \cos\theta + \Kone =0. \label{eq:definesthepolywithks}
\end{equation}
Expressions for the coefficients $\Kss$, $\Kcc$, $\Kcs$, $\Ks$, $\Kc$
and $\Kone$ are listed later in equations (\ref{eq:firstkdef}) to
(\ref{eq:lastkdef}).  We note that the left hand side of
equation~(\ref{eq:definesthepolywithks}) viewed as a function of
$\theta$, (i) is bounded, (ii) is real, (iii) is continuous with
period $2\pi$, (iv) is not constant (except in degenerate cases which
we will not consider), (v) has no Fourier components with period
smaller than $\pi$, and therefore (again excluding degenerate cases which an implementation would need to deal with)
has either two real roots or four real roots.

\par

 One method whereby
$\theta$ may be determined from
equation~(\ref{eq:definesthepolywithks}) is to replace $\cos\theta$
with $\pm\sqrt{1-\sin^2\theta}$ before then taking an appropriate
square in order to remove the $\sqrt{\cdots}$ resulting in a quartic
polynomial in $\sin\theta$ of the form shown later in
equation~(\ref{eq:quarticins}) where, for simplicity, $\sin\theta$ has
been abbreviated as $s$.  One must remember that in promoting
equation~(\ref{eq:definesthepolywithks}) to a quartic we have
introduced spurious solutions -- effectively those that have the
``wrong'' sign for $\cos\theta$, i.e.\ a sign which is incompatible
with equation~(\ref{eq:definesthepolywithks}).  Nevertheless, given
any solution $s=s_0$ of equation~(\ref{eq:quarticins}) one can
determine which sign of $\cos\theta$ is appropriate by returning to
the un-squared form and checking consistency.  For completeness, the
exact nature of the test required to determine the sign of
$\cos\theta$ is listed later in
equations~(\ref{eq:signspecificationbegins}) to
(\ref{eq:signspecificationends}).

\par

The two (or four) real roots $s\in\{s_1,s_2\}$ (or
$s\in\{s_1,s_2,s_3,s_4\}$) of (\ref{eq:quarticins}) may be obtained
analytically and non-iteratively by many methods (such as that of Ferrari) and thence candidate
values for $\sin\theta$ and $\cos\theta$ may be found by the methods
already described.  It remains only (i) to determine the magnitudes
$\left|{\bf p}\right|$ and $\left|{\bf q}\right|$ in terms of these
candidates, (ii) to dismiss any solutions yielding unphysical answers
(such as complex $\theta$ or negative $\left|{\bf p}\right|$ or
negative $\left|{\bf q}\right|$) and (iii) to determine which of the remaining solutions (if more than one) leads to the smallest value of either side of equation~(\ref{eq:balconfigsdfsdf}), this then being the desired result, namely $M_{T2}^2$. Steps (i) and (ii) may be achieved by noting that equation (\ref{eq:mombalone}) uniquely fixes the ratio $\poverq=\left|{\bf p}\right|/\left|{\bf q}\right|$ in terms of known quantities (see equation (\ref{eq:powverqfixed})), while the absolute value of $\left|{\bf p}\right|$ and $\left|{\bf q}\right|$ is fixed by taking the scalar product of equation (\ref{eq:splittingcondpoiy}) with ${{\slashed{\bf p}} }$ resulting in $\left|{\bf p}\right|=\magPtmiss K \poverq$ and  $\left|{\bf q}\right|=\magPtmiss K $ for $K$ as defined in (\ref{eq:kdefined}).

\par

The outcome is the following result for $M_{T2}$.
\begin{equation}\label{eq:ssdfwefgf}
\left.\left(M_{T2}(a^\mu, b^\mu,\slashed {\bf p})\right)^2\right|_{m_a=m_b=\chi=0} =
\begin{cases}
0 & \text{if $\eap  \eahbh \ge 0$ and  $ \ebp  \eahbh \le 0$}\\
2 {|{\bf a}|} \magPtmiss \thing \poverq (1 - \cosTheta  \ahdotbh - \sinTheta  \eahbh) & \text{\qquad otherwise,} 
\end{cases}
\end{equation}
where\footnote{Note that in the second line of \eqref{eq:ssdfwefgf} one could use $2 {|{\bf b}|} \magPtmiss \thing (1 - \cosTheta  \ahdotbh + \sinTheta  \eahbh)$ in place of $2 {|{\bf a}|} \magPtmiss \thing \poverq (1 - \cosTheta  \ahdotbh - \sinTheta  \eahbh)$.}
\begin{eqnarray}
\thing &=& \left[\poverq\left(\cosTheta  \bhdotph - \sinTheta  \ebp\right) + \cosTheta\ahdotph - \sinTheta\eap\right]^{-1}, \label{eq:kdefined}\\
\poverq &=& -\frac{
	  \cosTheta  \eap + \sinTheta  \ahdotph  }{
	  \cosTheta  \ebp + \sinTheta  \bhdotph  },\label{eq:powverqfixed}
\end{eqnarray}
in which $\sinTheta$ and $\cosTheta$ are defined by
\begin{eqnarray}
\sinTheta &=& s, \label{eq:signspecificationbegins}\\
\cosTheta &=&
\begin{cases}
+\sqrt{1-s^2 } & \text{if $\LHS-\RHS = 0$} \\
-\sqrt{1-s^2 } & \text{if $\LHS+\RHS = 0$}
\end{cases},
\end{eqnarray}
where
\begin{eqnarray}
\LHS &=&  \Kone + \Kcc + s (\Ks + (-\Kcc + \Kss) s),\\
\RHS &=& -(\Kc + \Kcs s)  \sqrt{1-s^2 }, \label{eq:signspecificationends}
\end{eqnarray}
and in which $s$ is the appropriate ``real, $K>0$'' root\footnote{The quartic polynomial in $s$ has four roots, of which at least two are real and at most two form a complex conjugate pair.  If there are only two real roots, one will lead to $K>0$ and the other to $K<0$.  If there is more than one real root having $K>0$, the one leading to the smallest value of $\mttwo$ should be chosen.  Degenerate cases are not discussed.} of the equation
\begin{equation}
\AAA s^4 + \BB s^3 + \CC s^2 + \DD s + \EE = 0\label{eq:quarticins}
\end{equation}
in which
\begin{eqnarray}
\AAA &=& \Kcs^2 + (\Kss - \Kcc)^2 \\
\BB &=&  2 (\Kcs \Kc + \Ks (\Kss - \Kcc))\\
\CC &=& \Ks^2 -\Kcs^2 + \Kc^2 + 2 (\Kss - \Kcc) (\Kone + \Kcc)\\
\DD &=&  2 (-\Kcs \Kc + \Ks (\Kone + \Kcc))\\
\EE &=& (\Kone - \Kc + \Kcc) (\Kone + \Kc + \Kcc)
\end{eqnarray}
wherein
\begin{eqnarray}
 \Kss &=& - \deltadotp  \eahbh \label{eq:firstkdef}\\
     \Kcc &=&  - \esigmap  \ahdotbh \\
     \Ks &=& \sigmadotp \\
     \Kc &=& \esigmap \\
     \Kcs &=& -\sigmadotp  \ahdotbh - \edeltap  \eahbh \\
     \Kone &=& 0  \label{eq:lastkdef}
\end{eqnarray}
in which we have introduced two new two-vectors $\bf \sigma$ and $\bf \delta$ according to
\begin{eqnarray}
\sigma &=& {\bf a} + {\bf b}, \qquad\text{and} \\
\delta &=& {\bf a} - {\bf b}.
\end{eqnarray}
Throughout the above we have adopted a notation in which unit-vectors carry a ``hat'', while a ``prime'' (as in ${\bf b}'$) indicates rotation of the two-vector through 90 degrees in the transverse plane.  A consequence of this is that
\begin{eqnarray}
({\bf v}.{\bf w}) &\equiv& \left|{\bf v}\right|\left|{\bf w}\right|\cos{\theta_{vw}}\qquad\text{and}\\
({\bf v}.{{\bf w}'}) &\equiv& \left|{\bf v}\right|\left|{\bf w}\right|\sin{\theta_{vw}}
\end{eqnarray}
for arbitrary two vectors ${\bf v}$ and ${\bf w}$.

\subsection{Remarks on the fully massless case ($a^2=b^2=\chi^2=0$).}

\label{sec:interpret}

We note that for real constants $\lambda$ and $\mu$ satisfying $\lambda\mu\ge0$, the solution in the above special case has the following property\footnote{The property described may easily be proved without using the $M_{T2}$ solution.  It is sufficient to note that the balanced condition (that the transverse masses on each sides are equal for the splitting that achieves the minimal transverse mass for either side) is not disturbed by scaling the missing momenta or by scaling both visible momenta equally.}
$$
\left.M_{T2}^2\left(\lambda a^\mu,\lambda b^\mu,\mu \slashed {\bf p}\right)\right|_{m_a=m_b=\chi=0}
= \lambda \mu
\left.M_{T2}^2\left(a^\mu,b^\mu,\slashed {\bf p}\right)\right|_{m_a=m_b=\chi=0}
$$
or equivalently
$$
\left.M_{T2}^2\left(a^\mu,b^\mu,\slashed{\bf p}\right)\right|_{m_a=m_b=\chi=0}
= \sqrt{\left|{\bf a}\right|\left|{\bf b}\right|}\left|{\slashed{\bf p}}\right|
\left.M_{T2}^2\left(\sqrt{\frac{\left|{\bf a}\right|}{\left|{\bf b}\right|}}\hat{\bf a},\sqrt\frac{\left|{\bf b}\right|}{\left|{\bf a}\right|}\hat{\bf b},\hat{\slashed{\bf p}}\right)\right|_{m_a=m_b=\chi=0}
$$
which could be interpreted as saying that the non-trivial dependence of $M_{T2}$ on its inputs in this special case is confined to three dimensionless parameters, of which two are relative angles of the visible and missing transverse momenta, and one is the ratio of the momenta of the two visible particles.

Another interpretation of this result, is that $\mttwo$ in the fully massless case can be decomposed into a ``magnitude'' part
$$
\rho_{T2} = \sqrt{\sqrt{\left|{\bf a}\right|\left|{\bf b}\right|}\left|{\slashed{\bf p}}\right|}
$$
 and an ``angular'' part
$$
\theta_{T2} =  M_{T2}\left(\sqrt{\frac{\left|{\bf a}\right|}{\left|{\bf b}\right|}}\hat{\bf a},\sqrt\frac{\left|{\bf b}\right|}{\left|{\bf a}\right|}\hat{\bf b},\hat{\slashed{\bf p}}\right)
$$ such that $\mttwo = \rho_{T2} \theta_{T2}$.  Interestingly, $\theta_{T2}$ seems to have only very mild dependence on ${\left|{\bf b}\right|}/{\left|{\bf a}\right|}$ and so is, to a relatively good approximation, just a universal function of $\theta_{a\slashed p}$ and $\theta_{b\slashed p}$ which is small for back to back events and large for pencil-like collimated events, multiplied by a normalisation function $f({\left|{\bf b}\right|}/{\left|{\bf a}\right|})$ whose maximum occurs when ${\left|{\bf b}\right|}/{\left|{\bf a}\right|}=1$ and which is only slowly varying with ${\left|{\bf b}\right|}/{\left|{\bf a}\right|}$. 

Accordingly, much of the ``signal'' structure of $\mttwo$ in the massless case comes from the $\rho_{T2}$ part. Finally, we remark in passing, that $\rho_{T2}$ is, in effect, a geometric mean of the magnitudes of the input momenta.  Contrast this with the effective mass $M_\mathrm{eff}$ \cite{Tovey:2000wk} (sometimes also referred to as $H_T$ or similar) which is proportional\footnote{Note that where the number of ingredients $n$ for $M_\mathrm{eff}$ and $H_T$ can vary between events, the constant of proportionality, $n$, will also vary between events.} to the algebraic mean of the input momenta.  We therefore learn that there is a sense in which $\mttwo$ is acting a bit like $M_\mathrm{eff}$ or $H_T$ but in ``log space'' rather than in ``linear space''.  One might conjecture whether there is anything to learn from that in the wider context ... for example, why is so much attention paid to linear sums?  It is an interesting open question as to whether it would be useful to construct geometric versions of the effective mass or $H_T$ such as
$$M_\mathrm{eff}^\mathrm{geom} = \left(\magPtmiss\prod_{i=1}^n |{\bf a}_i|\right)^{1/(n+1)}$$ (assuming one invisible) or
$$M_\mathrm{eff}^\mathrm{geom} = \left(\magPtmiss^2\prod_{i=1}^n |{\bf a}_i|\right)^{1/(n+2)}$$ (assuming two invisibles) or
$$H_T^\mathrm{geom} = \left( \prod_{i=1}^n |{\bf a}_i|\right)^{1/n}, $$ or variants thereof in which the mean was taken over the number of ``parents'' (1 or 2) rather than the number of constituents $n$.

\section{$M_{T2}$ in the ``$\slashed{\bf p}=Q({\bf a}+{\bf b})$'' case.}

\label{sec:funnycases}

In this section we consider results for $M_{T2}$ that are valid in the regime in which the missing momentum is proportional to (though not necessarily in the same direction as) the sum of the visible momenta from each side of the event.  In this section, masses are general and need not be zero.  In other words, we concern ourselves here with the case $\slashed{\bf p}=Q({\bf a}+{\bf b})$ for some real constant $Q$ satisfying $-\infty < Q < +\infty$. The results we will find are
\begin{itemize}\item
the general solution for $Q=-1$,\item the general solution for $Q=0$ and \item the general solution for any $Q$ but with the requirement that $m_a=m_b$ (=``$m$'').\end{itemize}
Before establishing our new results, we first comment on what is already known in these regimes. 
\subsection{Previous results}
The ``$Q=-1$ case'' corresponds to an absence of Upstream Transverse Momentum (UTM).  It has already been shown in \cite{Lester:2007fq} and \cite{Cho:2007dh} that in this case
\begin{align}
M_{T2}^{\rm bal}(a^\mu,b^\mu,\slashed{\bf p}=-({\bf a}+{\bf b}))^2  = \rhsOne. \label{eq:RHS1a}
\end{align}
One thing this shows is is that the dynamic dependence of $M_{T2}$ on its inputs (in that special case) is contained entirely within the contralinear boost invariant quantity $A_T$. The existing proofs provide no clear reason as to where that invariance comes from.\footnote{The dependence of (\ref{eq:RHS1a}) on $A_T$ was not evident in the labyrinthine result first published in \cite{Lester:2007fq}.  The exclusive dependence of the result on $A_T$ was first noted by \cite{Cho:2007dh}, ostensibly by simplification of the result of \cite{Lester:2007fq}. The contralinear boost invariance of the result only becomes manifest in the final step of that simplification, and thus provides little insight as to where it comes from.}  Herein we will re-prove that result using a method that maintains manifest contralinear boost invariance at all times, and in doing so (1) we will gain some insight as to where the invariance comes from, and (2) we will be lead to make further generalisations of the result to the case $Q\ne -1$.
\par
The $Q=0$ case (i.e. the case in which $\slashed{\bf p}_T=0$) received a small amount of attention in \cite{Barr:2003rg}.  Specifically, it was recorded therein that 
$$\left.{M_{T2}^{\rm bal}(a^\mu,b^\mu,\slashed{\bf p}=0)}\right|^2_{m_a=m_b=m}  
= \chi^2 + m^2 + \chi \sqrt{2(A_T + m^2)}$$
Herein we go beyond that result by generalising it to the case that $m_a\ne m_b$.  Furthermore we gain the result by a method maintaining manifest contralinear boost invariance throughout.

\par

The case where {\em both} $Q=+1$ {\em and} $m_a=m_b=m$ received, perhaps inadvertently, some attention in \cite{Cho:2009ve}.  Specifically \cite{Cho:2009ve} gave an expression for $M_{CT2}$ (note, not $M_{T2}$) in the case where the visible particles are massless ($m_a=m_b=0$) and there is no UTM (i.e.~$\slashed{\bf p}=-({\bf a}+{\bf b})$).  No explicit claims relating $M_{T2}$ solutions to $M_{CT2}$ solutions are made in \cite{Cho:2009ve}, however using (\ref{eq:mt2ismct2backwards}) we can see that the result of \cite{Cho:2009ve} corresponds to an expression for $M_{T2}$ in which the visible particles are still massless ($m_a=m_b=0$) but in which there is {\em a large amount} of UTM, since $\slashed{\bf p}=+({\bf a}+{\bf b})$.  Indeed, the observation that the $M_{CT2}$ result of \cite{Cho:2009ve} corresponded to an $M_{T2}$ result was the trigger for writing this paper.  We will re-prove that result, but our result will then go beyond it as it will neither require $Q=+1$, nor require the visible particles to me massless.

\subsection{The new results}
We shall prove the following results by methods that maintain manifest contralinear boost invariance at all times:
\begin{flalign} 
& M_{T2}^{\rm bal}(a^\mu,b^\mu,\slashed{\bf p}=0)^2 = \nonumber \\ 
 & = {\rhsForZeroPtmissPartOneAlternative  + \rhsForZeroPtmissPartTwoAlternative \label{eq:mt2ptmisszero} }\\
& \equiv { \rhsForZeroPtmissPartOne + \rhsForZeroPtmissPartTwo\label{eq:mt2ptmisszeroalt}}
\end{flalign}
and
\begin{eqnarray}
\lefteqn{   {M_{ T2}^{\rm bal} (a^\mu  ,  b^\mu,\slashed{\bf p}=+Q({\bf a}+{\bf b}))}^2 = } \nonumber \\
& = &       \rhsThree      \label{eq:mt2withqs}\\
&\equiv&\rhsThreeAlt\label{eq:mt2withqsalt}
\end{eqnarray}
\hide{
\comCL{check consistency with the earlier results quoted in this paper at fixed $Q$.}
\comCL{Note that I have not explicitly checked that an alternative $-$ sign in front of the square root (which would also satisfy the Euler-Lagrange equations) is {\em always} unphysical.}

\comCL{What about unbalanced cases? What about the collinear zero case?}
\comCL{Can anyone find an answer which works when $m_a \ne m_b$ ?}
}


 We note that the RHS of (\ref{eq:mt2ptmisszeroalt}) is always less than the RHS of (\ref{eq:RHS1a}). \hide{\comCL{Make something of this?}}

\subsubsection{Proof for the case when $\slashed{\bf p}=0$.}
\label{sec:pt2ptmissero}
A consequence of  $\slashed{\bf p}=0$ is that the invisible daughter
particles (being the only sources of missing transverse momentum) must
be back-to-back in the lab frame.
We can enforce the conditions (a) that the invisible daughter hypotheses be back-to-back, and (b) that they share a common mass, by writing $q=\bar p$. To calculate the value of $M_{T2}$ in the ``balanced'' case \hide{\comCL{check the ptmiss between a and b case at zero mass}} it is therefore sufficient to perform an Euler-Lagrange minimisation using the Lagrangian
\begin{equation}
{\mathcal L}=\frac{1}{2} \left((a+p)^2+(\bar b+p)^2\right)+\frac{\lambda}{2}\left(p^2-\chi^2\right)+\frac{\mu}{2}\left((a+p)^2-(\bar b +p)^2\right)
\end{equation}
in which $\lambda$ and $\mu$ are Lagrange multipliers, the former enforcing the constraint that the invisible daughters have mass $\chi$, and the latter enforcing the constraint which gives us the balanced case.  The resultant Euler-Lagrange equation for $p$ (i.e.~${\partial \mathcal L}/{\partial p}=0$) then reduces to
\begin{equation}
p = A a + B \bar b \label{eq:linpincase2} 
\end{equation}
for unknown constants $A$ and $B$ (functions of the Lagrange multipliers).  We can determine $A$ and $B$ by substituting them back into the two constraints, making $A$ and $B$ the solution of the simultaneous equations
\begin{eqnarray}
(Aa+B\bar b)^2&=&\chi^2 \\
\left((A+1)a + B \bar b\right)^2 &=& \left(A a + (B+1)\bar b\right)^2
\end{eqnarray}
which reduce to 
\begin{eqnarray}
A^2 m_a^2 + B^2 m_b^2 + 2 A B A_T &=& \chi^2, \\
(2A+1) m_a^2 - (2B+1) m_b^2 &=& 2(A-B) A_T,
\end{eqnarray}
where we have once again defined $A_T=(a.\bar b)$.
It now only remains to solve these two simultaneous equations in order to determine $A$ and $B$ in terms of $A_T$, $\chi$, $m_a$ and $m_b$, and then to substitute the values so determined into equation~(\ref{eq:linpincase2}) to determine $p$, before finally to substituting this value of $p$ into $(a+p)^2$ (or $(\bar b+p)^2$ since it will be the same) in order to determine $M_{T2}^2$ for this balanced case.  This leads to the result shown earlier in equations~(\ref{eq:mt2ptmisszero}) and (\ref{eq:mt2ptmisszeroalt}) and concludes the proof.

%

\subsubsection{Proof for the case when $\slashed{\bf p}\ne 0$.}

We begin by defining two new transverse Lorentz vectors $k$ and $r$ according to $k=Qa-p$ and $r=Qb-q$.  
Next we demonstrate that if $k = \bar r$ then (i) the missing momentum condition $\slashed{\bf p}={\bf p}+{\bf q}$, and (ii) the condition for the critical $M_{T2}$ splitting hypothesis to be {\em balanced}, are both satisfied (provided that $m_a=m_b$ or $Q= -1$).  Let us consider (i) first.  If $k=\bar r$ then $Qa-p = Q\bar b - \bar q$ which implies $Q(a-\bar b) = p - \bar q$ which, taking the transverse components, implies $Q({\bf a}+{\bf b})={\bf p}+{\bf q}$ as required.  Now we must prove (ii).  $k=\bar r$ implies $k^2=r^2$ which implies $(Qa-p)^2=(Qb-q)^2$ which implies $Q^2 m_a^2 -2Q(a.p) + \chi^2 = Q^2m_b^2 -2Q(b.q) + \chi^2$ which (if $Q\ne0$)\footnote{We have already considered the $Q=0$ case separately and with greater generality (see for example equation~(\ref{eq:mt2ptmisszero}) valid for $m_a\ne m_b$) and therefore the invalidity of the proof in the case $Q= 0$ need not concern us here.  However, for completeness we note that the limit $|Q|\rightarrow 0$ of the solution we are about to obtain is well defined and is the same as that of (\ref{eq:mt2ptmisszero}), at least in the case $m_a=m_b$ under consideration, and therefore the answer need not carry $Q\ne0$ qualifiers. \hide{\comCL{check!}}} implies $2(a.p)-2(b.q)=Q(m_a^2-m_b^2)$ which implies $(a+p)^2-(b+q)^2 =m_a^2-m_b^2 + Q(m_a^2-mb^2) = (1+Q)(m_a^2-m_b^2)$.  This allows us to see, as required, that $k=\bar r$ implies that the ``balanced'' condition is satisfied if $Q=-1$ or $m_a=m_b$.

\par

We are now in a position to claim that $M_{T2}$ for the case under consideration will be given by the solution to the Euler Lagrange problem with free parameters $k$, $\lambda$ and $\mu$ with Lagrangian
\begin{eqnarray}
{\mathcal L}(k,\lambda,\mu) &=& (a+p)^2 + \frac{\lambda}{2}(p^2 -\chi^2)+ \frac{\mu}{2}(q^2 -\chi^2)\\
&=& ((1+Q)a-k)^2 + \frac{\lambda}{2}((Qa-k)^2 -\chi^2)+ \frac{\mu}{2}((Q\bar b -k)^2 -\chi^2)
\end{eqnarray}
which gives us again a (different) Euler-Lagrange equation for $k$ of the form
$$k=A a + B \bar b\label{eq:kincasetoday}$$
for some, as yet undetermined, constants $A$ and $B$ which may be found by solving the remaining constraint equations associated with $\lambda$ and $\mu$ namely:
\begin{eqnarray}
((A-Q)a + B \bar b)^2&=&\chi^2 \\
((A a + ( B-Q) \bar b)^2&=&\chi^2 
\end{eqnarray}
or equivalently
\begin{eqnarray}
(A -Q)^2 m_a^2 + B^2 m_b^2 + 2 (A -Q) B A_T &=&\chi^2 \\
A^2 m_a^2 + (B - Q)^2 m_b^2 + 2 A (B -Q) A_T &=&\chi^2.
\end{eqnarray}
\hide{\comCL{Note that such a soln, valid on two ``axes'' (mass difference at Q=-1 and Q at massdifference=0) can perhaps be analytically continued onto the whole Q.massdifffernece plane? No!}}
Taking the difference we discover $$2 AQ(A_T-m_a^2)+ Q^2 m_a^2=2BQ(A_T -m_b^2)+ Q^2m_b^2$$ which (if $Q\ne 0$ as before) allows us to eliminate either A or B from the preceding equations, leaving at worst a quadratic expression for whichever quantity remains.  With $A$ and $B$ now determined in terms of $Q$, $m_a^2$, $m_b^2$ and $A_T$ it only remains to find the balanced $M_{T2}^2$ solution by substituting into the expression $M^2 = \frac{1}{2}\left((a+p)^2+(b+q)^2\right) = \frac{1}{2}\left((a+Qa-k)^2+(\bar b+ Q\bar b-k)^2\right)$ for $k$ as defined in equation~(\ref{eq:kincasetoday}).  This results in the single expression:
\begin{equation}
M^2=\singleExpression
\end{equation}
which we recall is is only ``meaningful'' if either $Q=-1$ or $m_a = m_b$.  Specialising the above expression for $M^2$ for both of those cases leads to the right hand sides of (\ref{eq:mt2withqs}) and (\ref{eq:mt2withqsalt}) and thus concludes the proof.\label{sec:conjectures}

\begin{multicols}{2}

\section{Conclusions}
We have detailed non-iterative algorithms for calculating $\mttwo$ valid in a number of new special cases.  One of these is the ``fully massless case'' which is the scenario in which $\mttwo$ is used most frequently at the LHC.  The other cases (most but not all of which are new) apply when the transverse missing momentum is parallel or anti-parallel to the vector sum of the visible momenta.  Furthermore, in the cases for which non-iterative solutions were already known, we have found new derivations which are manifestly contralinear boost invariance at all times, providing advances in insight over earlier derivations.  Along the way, we have stumbled in Section~\ref{sec:conjectures} across a number of interesting conjectures into the nature of variables like $M_\mathrm{eff}$ and $H_T$, and have also gained therein better insight into the nature of $\mttwo$ as a geometric mean in the fully massless case.

\section{Acknowledgements}
The author would like to thank Alan Barr, Hsin-Chia Cheng, Sky French, James Frost, Zhenyu Han, Teng-Jian Khoo, Colin Lally, Tanya Sandoval, Dan Tovey, and especially Chris Young (whose observation that $\mttwo(a^\mu,b^\mu,\slashed{p}) = M_{CT2}(a^\mu,b^\mu,-\slashed{p})$ prompted the writing of the first draft) for helpful discussions and encouragement.
\appendix

\section{Appendix}

  
Many techniques have been proposed for measuring the masses of the new
particles which it is hoped the Large Hadron Collider will produce
(see \cite{Barr:2010zj} for a recent review). Some of these techniques
use the kinematic variable known as $\mttwo$
\cite{Lester:1999tx} which may be thought of either as a
natural extension of the transverse mass $M_T$
\cite{Arnison:1983rp,*Banner:1983jy} to events containing pairs of
mother particles, each undergoing a decay into a mixture of visible
and invisible daughter particles, or as an event-by-event bound on the
kinematic properties of such events \cite{Cheng:2008hk}.

\par

Much of the literature that has developed $\mttwo$ methods
\cite{Allanach:2000kt,Barr:2002ex,Barr:2003rg,Lester:2007fq,Cho:2007qv,Gripaios:2007is,Barr:2007hy,Cho:2007dh,Ross:2007rm,Nojiri:2007pq,Tovey:2008ui,Cho:2008cu,Serna:2008zk,Barr:2008ba,Cho:2008tj,Burns:2008va,Barr:2008hv,Barr:2009mx,Barr:2009jv,Polesello:2009rn,Kim:2009si,Konar:2009wn,Konar:2009qr}
is concerned with the kinematic properties of the variable, the
properties of its endpoints, and how (or whether) one can use these to
place constraints on, or perhaps even measure\footnote{Thus far,
$\mttwo$ has only been used once in anger to measure the mass of a
particle -- the top-quark in CDF in the dilepton channel
\cite{Aaltonen:2009rm}. The results are promising, and we are told
that the top quark mass measurement with $\mttwo$ has ``the smallest
total systematic uncertainty'' of any in that channel
\cite{Aaltonen:2009rm}.}, the masses of new pair produced particles
and/or their invisible daughters.  An entirely different use for
$\mttwo$ has been highlighted \cite{Barr:2009wu,Barr:2010ii} by
members one of the large general-purpose LHC experiments --
identifying properties of $\mttwo$ that explain why it
is useful as a ``cut'' or ``discovery'' variable.  As a consequence of its sensitivity to the mass scale of the pair produced parents, and as a consequence of its definition as a kinematic bound, it is particularly good at suppressing the low multiplicity low-mass-scale standard model
processes (principally QCD and pair production of top quarks) which can be backgrounds to
new-physics signatures with few visible
particles into the final state.\footnote{Consider, for example,
supersymmetry in the case that the only thing that can be produced are
squark pairs, each decaying to a quark jet and an (invisible)
neutralino.}  Early indications of the performance of $\mttwo$ in
early ATLAS data were very encouraging \cite{Collaboration:1273174},
and indeed it was pleasing to see that the most stringent expected limits on
di-squark production from the 2010 LHC data came came from the
use, by ATLAS \cite{Collaboration:2011qk}, of $\mttwo$ in this way.

\par



However, as the instantaneous luminosity increases, it becomes necessary
either to pre-scale triggers\footnote{To ``pre-scale'' a trigger means
to accept at random only a fixed and pre-determined {\rm fraction} of
the events that pass it.} or to increase the trigger thresholds
(e.g.~the minimum transverse jet momenta).  In particular, the QCD
dijet cross section is so large that long before design luminosity is
reached, one will find it necessary to apply to single and dijet
triggers either very large pre-scales or very high jet $p_T$ thresholds
to prevent QCD events saturating the trigger.  All this is bad news
for any new-physics searches that hope to look for signals containing
only two jets in association with missing transverse momentum, such as supersymmetric
models in which all sparticles are heavy except the squarks and the
neutralino LSP (lightest supersymmetric particle).  Searches for such
``low multiplicity'' signals are compromised if the
only triggers they can pass are single or dijet or missing transverse
momentum triggers.  Fortunately this is not the end of the story.
Since the majority of the QCD events contain back-to-back jets, some
experiments have
implemented ``$\Delta\Phi$'' triggers -- i.e.~triggers which only accept events
if the leading two jets (above some $p_T$ threshold) have an angular
separation in the transverse plane which is less than a pre-defined
value (such as $0.9\pi$).  QCD events find it much harder to pass $\Delta\Phi$ triggers than,
say, di-squark susy events, and so such triggers can remain
un-pre-scaled for a greater length of time than the corresponding mono-
and di-jet triggers.  Though $\Delta\Phi$ triggers are conceptually easy to
understand and implement, $\mttwo$ is expected to discriminate QCD
from susy much better than $\Delta\Phi$ \cite{Randall:2008rw}.  The lack of fast methods for evaluating $\mttwo$ has, however, presented a hurdle to the adoption of $\mttwo$ as a trigger. \hide{ \footnote{According to google, ATLAS appears to be developing an MCT trigger \cite{atlasmcttrigger1}.}}

As it is not possible to write down closed-form analytic expressions
for $\mttwo$ in the general case,\footnote{It is possible to write
$\mttwo$ as a real root of a number of different polynomials, but
sadly all such polynomials have been found to have degree greater than
four (one is documented in \cite{Cheng:2008hk}, and others are known
to the author) and so it seems unlikely that a closed-form analytic
expression for the variable exists. An unclaimed \pounds 200 prize
awaits the provider of a counter example to this suggestion.}
$\mttwo$ is usually evaluated using numerical libraries such as
\cite{oxbridgeStransverseMassLibrary} and
\cite{zenuhanStransverseMassLibrary} which use iterative algorithms and are therefore too slow to use in LHC experiment triggers.\footnote{All the present $\mttwo$
algorithms are iterative, e.g.~through a dependence on numerical
minimisation algorithms, or through use of the ``bisection algorithm''
of \cite{Cheng:2008hk}.}

One of the motivations for this study is therefore the hope that methods of calculating $\mttwo$ quickly and reliably can be found, in the cases of interest to experiments, so that $\mttwo$ may be implement as a trigger variable.

The other arguably more important motivation for this study is pure mathematical interest.  Ref \cite{Cheng:2008hk} uncovered very useful mathematical insights into the nature of $\mttwo$
which allowed the creation of what is, at present, the fastest and
most accurate algorithm for the evaluation of $\mttwo$.\footnote{The authors
of \cite{Cheng:2008hk} refer to this as the ``bisection algorithm''.
A very simple C++ implementation of the bisection algorithm,
consisting of a single ``{\tt .h}'' file and a single ``{\tt .c}''
file, may be downloaded from \cite{zenuhanStransverseMassLibrary}.
The same implementation is also distributed within
\cite{oxbridgeStransverseMassLibrary} which, though harder to use, may
be of interest to developers or persons comparing implementations of
$\mttwo$ and similar variables. }   It is not always possible to predict what fruit a mathematical investigation will bring.  The buds ripening here are those interpretations of Section~\ref{sec:interpret}.

\hide{Though $\mttwo$ may turn out to have an independent role as a
discovery variable in its own right \cite{Barr:2009wu}, it has
typically been used or viewed as an event variable which generates
(for each event) a lower bound on the mass of the pair-produced mother
particles presumed to be in the event, in terms of a hypothesised
value of the mass of the invisible daughter particles in the event.  As
such, the variable is often written ``$\mttwo(\chi)$'', emphasising
that it not merely a number, but is in fact a function of the
hypothesised daughter particle mass, $\chi$.\footnote{Though the
somewhat asymmetric ``first hypothesise a daughter mass, then obtain a
bound on the parent mass'' interpretation of $\mttwo$ is the
interpretation which has most commonly been used, it was noted in
\cite{Cheng:2008hk} that this asymmetric viewpoint is not fundamental.
It was shown in \cite{Cheng:2008hk} that the reverse inference
can also be drawn from $\mttwo$ ({\it i.e.}~that the inverse function
$\chi(\mttwo)$) describes an {\em upper} bound on the invisible {\em
daughter} particle mass in terms of hypothesised {\em parent} mass)
and that moreover the most general interpretation of the $\mttwo$
curve is that, for any particular event, the curve $\mttwo(\chi)$
delineates the {\em boundary} of the region of $(m_{\mbox{daughter}},
m_{\mbox{parent}})$-space which is kinematically consistent (or indeed
inconsistent) with the hypothesised event structure.  The consequences
of this are discussed in more detail in \cite{Barr:2009jv}.}
}

\bibliographystyle{JHEP-withSlacCitation}
\bibliography{mt2_special_cases}

\providecommand{\href}[2]{#2}\begingroup\raggedright\begin{thebibliography}{10}

\bibitem{Lester:1999tx}
C.~G. Lester and D.~J. Summers, {\it {Measuring masses of semiinvisibly
  decaying particles pair produced at hadron colliders}},  {\em Phys. Lett.}
  {\bf B463} (1999) 99--103,
  [\href{http://xxx.lanl.gov/abs/hep-ph/9906349}{{\tt hep-ph/9906349}}].

\bibitem{Barr:2003rg}
A.~Barr, C.~Lester, and P.~Stephens, {\it {m(T2) : The Truth behind the
  glamour}},  {\em J. Phys.} {\bf G29} (2003) 2343--2363,
  [\href{http://xxx.lanl.gov/abs/hep-ph/0304226}{{\tt hep-ph/0304226}}].

\bibitem{Lester:2007fq}
C.~Lester and A.~Barr, {\it {$M_{TGen}$} : Mass scale measurements in
  pair-production at colliders},  {\em JHEP} {\bf 12} (2007) 102,
  [\href{http://xxx.lanl.gov/abs/0708.1028}{{\tt arXiv:0708.1028}}].

\bibitem{Cho:2009ve}
W.~S. Cho, J.~E. Kim, and J.-H. Kim, {\it {Shining on buried new particles}},
  \href{http://xxx.lanl.gov/abs/0912.2354}{{\tt arXiv:0912.2354}}.

\bibitem{Cheng:2008hk}
H.-C. Cheng and Z.~Han, {\it Minimal kinematic constraints and {$M_{T2}$}},
  {\em JHEP} {\bf 12} (2008) 063,
  [\href{http://xxx.lanl.gov/abs/0810.5178}{{\tt arXiv:0810.5178}}].

\bibitem{Tovey:2008ui}
D.~R. Tovey, {\it {On measuring the masses of pair-produced semi-invisibly
  decaying particles at hadron colliders}},  {\em JHEP} {\bf 04} (2008) 034,
  [\href{http://xxx.lanl.gov/abs/0802.2879}{{\tt arXiv:0802.2879}}].

\bibitem{Barr:2009wu}
A.~J. Barr and C.~Gwenlan, {\it {The race for supersymmetry: using $M_{T2}$ for
  discovery}},  {\em Phys. Rev.} {\bf D80} (2009) 074007,
  [\href{http://xxx.lanl.gov/abs/0907.2713}{{\tt arXiv:0907.2713}}].

\bibitem{Barr:2010ii}
A.~J. Barr, C.~Gwenlan, C.~G. Lester, and C.~J.~S. Young, {\it {A comment on
  'Amplification of endpoint structure for new particle mass measurement at the
  LHC'}},  \href{http://xxx.lanl.gov/abs/1006.2568}{{\tt arXiv:1006.2568}}.

\bibitem{Collaboration:1273174}
{\bf {ATLAS}} Collaboration, {The ATLAS Collaboration}, {\it Early
  supersymmetry searches in channels with jets and missing transverse momentum
  with the {ATLAS} detector},  Tech. Rep. ATLAS-COM-CONF-2010-066, {CERN},
  Geneva, Jun, 2010.

\bibitem{Collaboration:2011qk}
{\bf ATLAS} Collaboration, {\it {Search for squarks and gluinos using final
  states with jets and missing transverse momentum with the ATLAS detector in
  sqrt(s) = 7 TeV proton-proton collisions}},
  \href{http://xxx.lanl.gov/abs/1102.5290}{{\tt arXiv:1102.5290}}.

\bibitem{Cho:2008cu}
W.~S. Cho, K.~Choi, Y.~G. Kim, and C.~B. Park, {\it {Measuring the top quark
  mass with $m_{T2}$ at the {LHC}}},  {\em Phys. Rev.} {\bf D78} (2008) 034019,
  [\href{http://xxx.lanl.gov/abs/0804.2185}{{\tt arXiv:0804.2185}}].

\bibitem{Aaltonen:2009rm}
{\bf CDF} Collaboration, T.~Aaltonen {\em et.~al.}, {\it {Top Quark Mass
  Measurement using mT2 in the Dilepton Channel at CDF}},  {\em Phys. Rev.}
  {\bf D81} (2010) 031102, [\href{http://xxx.lanl.gov/abs/0911.2956}{{\tt
  arXiv:0911.2956}}].

\bibitem{Tovey:2000wk}
D.~R. Tovey, {\it {Measuring the {SUSY} mass scale at the {LHC}}},  {\em Phys.
  Lett.} {\bf B498} (2001) 1--10,
  [\href{http://xxx.lanl.gov/abs/hep-ph/0006276}{{\tt hep-ph/0006276}}].

\bibitem{Cho:2007dh}
W.~S. Cho, K.~Choi, Y.~G. Kim, and C.~B. Park, {\it {Measuring superparticle
  masses at hadron collider using the transverse mass kink}},  {\em JHEP} {\bf
  02} (2008) 035, [\href{http://xxx.lanl.gov/abs/0711.4526}{{\tt
  arXiv:0711.4526}}].

\bibitem{Barr:2010zj}
A.~J. Barr and C.~G. Lester, {\it {A Review of the Mass Measurement Techniques
  proposed for the Large Hadron Collider}},  {\em J. Phys.} {\bf G37} (2010)
  123001, [\href{http://xxx.lanl.gov/abs/1004.2732}{{\tt arXiv:1004.2732}}].

\bibitem{Arnison:1983rp}
{\bf UA1} Collaboration, G.~Arnison {\em et.~al.}, {\it {Experimental
  observation of isolated large transverse energy electrons with associated
  missing energy at $s^{1/2}$ = 540 GeV}},  {\em Phys. Lett.} {\bf B122} (1983)
  103--116.

\bibitem{Allanach:2000kt}
B.~C. Allanach, C.~G. Lester, M.~A. Parker, and B.~R. Webber, {\it {Measuring
  sparticle masses in non-universal string inspired models at the {LHC}}},
  {\em JHEP} {\bf 09} (2000) 004,
  [\href{http://xxx.lanl.gov/abs/hep-ph/0007009}{{\tt hep-ph/0007009}}].

\bibitem{Barr:2002ex}
A.~J. Barr, C.~G. Lester, M.~A. Parker, B.~C. Allanach, and P.~Richardson, {\it
  {Discovering anomaly-mediated supersymmetry at the {LHC}}},  {\em JHEP} {\bf
  03} (2003) 045, [\href{http://xxx.lanl.gov/abs/hep-ph/0208214}{{\tt
  hep-ph/0208214}}].

\bibitem{Cho:2007qv}
W.~S. Cho, K.~Choi, Y.~G. Kim, and C.~B. Park, {\it Gluino stransverse mass},
  {\em Phys. Rev. Lett.} {\bf 100} (2008) 171801,
  [\href{http://xxx.lanl.gov/abs/0709.0288}{{\tt arXiv:0709.0288}}].

\bibitem{Gripaios:2007is}
B.~Gripaios, {\it Transverse observables and mass determination at hadron
  colliders},  {\em JHEP} {\bf 02} (2008) 053,
  [\href{http://xxx.lanl.gov/abs/0709.2740}{{\tt arXiv:0709.2740}}].

\bibitem{Barr:2007hy}
A.~J. Barr, B.~Gripaios, and C.~G. Lester, {\it Weighing {WIMPs} with kinks at
  colliders: Invisible particle mass measurements from endpoints},  {\em JHEP}
  {\bf 02} (2008) 014, [\href{http://xxx.lanl.gov/abs/0711.4008}{{\tt
  arXiv:0711.4008}}].

\bibitem{Ross:2007rm}
G.~G. Ross and M.~Serna, {\it Mass determination of new states at hadron
  colliders},  {\em Phys. Lett.} {\bf B665} (2008) 212--218,
  [\href{http://xxx.lanl.gov/abs/0712.0943}{{\tt arXiv:0712.0943}}].

\bibitem{Nojiri:2007pq}
M.~M. Nojiri, G.~Polesello, and D.~R. Tovey, {\it {A hybrid method for
  determining {SUSY} particle masses at the {LHC} with fully identified cascade
  decays}},  {\em JHEP} {\bf 05} (2008) 014,
  [\href{http://xxx.lanl.gov/abs/0712.2718}{{\tt arXiv:0712.2718}}].

\bibitem{Serna:2008zk}
M.~Serna, {\it {A short comparison between $m_{T2}$ and $m_{CT}$}},  {\em JHEP}
  {\bf 06} (2008) 004, [\href{http://xxx.lanl.gov/abs/0804.3344}{{\tt
  arXiv:0804.3344}}].

\bibitem{Barr:2008ba}
A.~J. Barr, G.~G. Ross, and M.~Serna, {\it The precision determination of
  invisible-particle masses at the {LHC}},  {\em Phys. Rev.} {\bf D78} (2008)
  056006, [\href{http://xxx.lanl.gov/abs/0806.3224}{{\tt arXiv:0806.3224}}].

\bibitem{Cho:2008tj}
W.~S. Cho, K.~Choi, Y.~G. Kim, and C.~B. Park, {\it {$M_{T2}$-assisted on-shell
  reconstruction of missing momenta and its application to spin measurement at
  the {LHC}}},  {\em Phys. Rev.} {\bf D79} (2009) 031701,
  [\href{http://xxx.lanl.gov/abs/0810.4853}{{\tt arXiv:0810.4853}}].

\bibitem{Burns:2008va}
M.~Burns, K.~Kong, K.~T. Matchev, and M.~Park, {\it Using subsystem $m_{T2}$
  for complete mass determinations in decay chains with missing energy at
  hadron colliders},  {\em JHEP} {\bf 03} (2009) 143,
  [\href{http://xxx.lanl.gov/abs/0810.5576}{{\tt arXiv:0810.5576}}].

\bibitem{Barr:2008hv}
A.~J. Barr, A.~Pinder, and M.~Serna, {\it {Precision Determination of
  Invisible-Particle Masses at the {CERN} {LHC}: II}},  {\em Phys. Rev.} {\bf
  D79} (2009) 074005, [\href{http://xxx.lanl.gov/abs/0811.2138}{{\tt
  arXiv:0811.2138}}].

\bibitem{Barr:2009mx}
A.~J. Barr, B.~Gripaios, and C.~G. Lester, {\it {Measuring the Higgs boson mass
  in dileptonic W-boson decays at hadron colliders}},  {\em JHEP} {\bf 07}
  (2009) 072, [\href{http://xxx.lanl.gov/abs/0902.4864}{{\tt
  arXiv:0902.4864}}].

\bibitem{Barr:2009jv}
A.~J. Barr, B.~Gripaios, and C.~G. Lester, {\it {Transverse masses and
  kinematic constraints: from the boundary to the crease}},  {\em JHEP} {\bf
  11} (2009) 096, [\href{http://xxx.lanl.gov/abs/0908.3779}{{\tt
  arXiv:0908.3779}}].

\bibitem{Polesello:2009rn}
G.~Polesello and D.~R. Tovey, {\it {Supersymmetric particle mass measurement
  with the boost-corrected contransverse mass}},  {\em JHEP} {\bf 03} (2010)
  030, [\href{http://xxx.lanl.gov/abs/0910.0174}{{\tt arXiv:0910.0174}}].

\bibitem{Kim:2009si}
I.-W. Kim, {\it Algebraic singularity method for mass measurement with missing
  energy},  {\em Phys. Rev. Lett.} {\bf 104} (2010) 081601,
  [\href{http://xxx.lanl.gov/abs/0910.1149}{{\tt arXiv:0910.1149}}].

\bibitem{Konar:2009wn}
P.~Konar, K.~Kong, K.~T. Matchev, and M.~Park, {\it {Superpartner mass
  measurements with 1D decomposed $M_{T2}$}},
  \href{http://xxx.lanl.gov/abs/0910.3679}{{\tt arXiv:0910.3679}}.

\bibitem{Konar:2009qr}
P.~Konar, K.~Kong, K.~T. Matchev, and M.~Park, {\it Dark matter particle
  spectroscopy at the {LHC}: Generalizing $m_{T2}$ to asymmetric event
  topologies},  \href{http://xxx.lanl.gov/abs/0911.4126}{{\tt
  arXiv:0911.4126}}.

\bibitem{Randall:2008rw}
L.~Randall and D.~Tucker-Smith, {\it {Dijet Searches for Supersymmetry at the
  {LHC}}},  {\em Phys. Rev. Lett.} {\bf 101} (2008) 221803,
  [\href{http://xxx.lanl.gov/abs/0806.1049}{{\tt arXiv:0806.1049}}].

\bibitem{oxbridgeStransverseMassLibrary}
A.~J. Barr and C.~G. Lester, ``Oxbridge stransverse mass library.''
\newblock \url{http://www.hep.phy.cam.ac.uk/~lester/mt2/index.html}.

\bibitem{zenuhanStransverseMassLibrary}
H.-C. Cheng and Z.~Han, ``{UCD} stransverse mass library.''
\newblock
  \url{http://particle.physics.ucdavis.edu/hefti/projects/doku.php?id=wimpmass%
}.

\end{thebibliography}\endgroup
\end{multicols}
\end{document}